\documentclass[showpacs,prd,nofootinbib,showkeys,twocolumn,10pt,aps,longbibliography]{revtex4-1}
\usepackage{hyperref} 
\usepackage{amsmath}
\usepackage{graphicx}
\usepackage[caption=false,captionskip=10pt]{subfig}
\usepackage[final]{feynmp}
\usepackage{tikz}
\usetikzlibrary{decorations.pathmorphing,decorations.markings,trees,arrows,patterns}
\usepackage{comment}
\usepackage{color}

\definecolor{darkgreen}{rgb}{0,.7,0}
\definecolor{linkblue}{rgb}{0.,0.,0.9333}

\hypersetup{
    pdffitwindow=true,      
    pdfauthor={W. A. Horowitz},     
    pdfnewwindow=true,      
    bookmarksnumbered=true,
    colorlinks=true,       
    linkcolor=red,          
    citecolor=darkgreen,        
    filecolor=magenta,      
    urlcolor=cyan,           
    menucolor=blue,
    baseurl=http://www.arxiv.org/abs/,
}

\newcommand{\fig}[1]{Fig.~\ref{fig:#1}}

\newcommand{\eq}[1]{Eq.~(\ref{eq:#1})}

\newcommand{\infinity}{\infty}

\begin{document}

\title{\texorpdfstring{A Complete Diagrammatic Implementation of the \\ Kinoshita-Lee-Nauenberg Theorem at Next-to-Leading Order}{A Complete Diagrammatic Implementation of the Kinoshita-Lee-Nauenberg Theorem at Next-to-Leading Order}}
\author{Abdullah Khalil}
\email[Email: ]{abdullah@aims.ac.za}
\affiliation{Department of Physics, University of Cape Town, Private Bag X3, Rondebosch 7701, South Africa}
\affiliation{Department of Physics, Cairo University, Giza 12613, Egypt}
\author{W.\ A.\ Horowitz}
\email[Email: ]{wa.horowitz@uct.ac.za}
\affiliation{Department of Physics, University of Cape Town, Private Bag X3, Rondebosch 7701, South Africa}

\begin{abstract}
We show for the first time in over 50 years how to correctly apply the Kinoshita-Lee-Nauenberg theorem diagrammatically in a next-to-leading order scattering process.  We improve on previous works by including all initial and final state soft radiative processes, including absorption and an infinite sum of partially disconnected amplitudes.  Crucially, we exploit the Monotone Convergence Theorem to prove that our delicate rearrangement of this formally divergent series is correct.  This rearrangement yields a factorization of the infinite contribution from the initial state soft photons that then cancels in the physically observable cross section.  We derive the first complete next-to-leading order, high-energy Rutherford elastic scattering cross section in the $\overline{\mathrm{MS}}$ renormalization scheme as an explicit example of our procedure. 
\end{abstract}

\maketitle
\tikzset{
    photon/.style={decorate, draw opacity=1, decoration={snake}, draw=black},
    particle/.style={draw=black, postaction={decorate},
        decoration={markings,mark=at position .5 with {\arrow[draw=black]{>}}}    }    ,
    antiparticle/.style={draw=black, postaction={decorate},
        decoration={markings,mark=at position .5 with {\arrow[draw=black]{<}}}},
    gluon/.style={decorate, draw=black,
        decoration={snake,amplitude=4pt, segment length=5pt}}
 }
\tikzset{
    pho/.style={draw opacity=0,decoration={markings,mark=at position .5 with {\arrow[draw=black,draw opacity=1,line width=0.5pt,scale=1.5]{<}}},
    postaction = {decorate}, 
    postaction = {draw = black, draw opacity = 1, decoration= {snake},decorate}
}}
\section{Introduction}
Infinities are ubiquitous in perturbative quantum field theoretic derivations of cross sections beyond leading order.  In order to make any beyond leading order result sensible, one must render these infinities harmless.  Renormalization eliminates ultraviolet (UV) divergences.  And the Kinoshita-Lee-Nauenberg (KLN) theorem \cite{Kinoshita:1962ur,Lee:1964is} assures that a summation over degenerate initial and final states removes all infrared (IR) divergences.  We show here for the first time a complete diagrammatic application of the KLN theorem at next-to-leading order accuracy.

Infrared divergences come in two flavors: \emph{soft}, due to the massless nature of the radiation (e.g.\ the massless photon in QED), and \emph{collinear}, which comes from treating the radiating particle as massless (e.g.\ the electron in QED).  Even before quantum field theory existed, Bloch and Nordsieck (BN) \cite{Bloch:1937pw} proved, in modern language, that the soft infrared divergences due to the massless radiated photon cancel when one sums over the degenerate final states of the system.  To compute an elastic Rutherford scattering cross section for an electron to scatter off an infinitely massive particle at next-to-leading order (NLO), invoking the BN theorem implies including the $1\rightarrow2$ process at leading order (LO) with the $1\rightarrow1$ process at NLO so long as the photon emitted in the $1\rightarrow2$ process possesses an undetectably small energy $E_\gamma<\Delta$.

One is interested in the massless radiator limit of a theory because at NLO and beyond, the coupling is multiplied by a logarithm of the scattered particle's mass; for asymptotically large energies, the finite coupling times the logarithm becomes larger than 1, and the perturbative expansion is uncontrolled.  

25 years after BN, Kinoshita \cite{Kinoshita:1962ur} and Lee and Nauenberg \cite{Lee:1964is} proved that the collinear infrared divergences due to a massless radiating particle cancel when one sums over both degenerate final and, crucially, initial states.  In their paper, Lee and Nauenberg (LN) included an example demonstrating the infrared finite cross section for elastic Rutherford scattering computed at NLO when one integrates over a degenerate final-state energy window and degenerate initial and final state angular windows.

Since the publication of the KLN papers in the early 1960's, there has been confusion regarding the proper treatment of the degenerate, \emph{soft} initial state radiation.  LN correctly identified the need to include \emph{partially disconnected} amplitudes in which a flying radiation does not interact with any other part of the amplitude.  These partially disconnected amplitudes can interfere with fully connected amplitudes in which a radiation is both absorbed and emitted by the radiator, forming a connected cut diagram.  However, LN did not include degenerate initial state soft photons in their treatment.  Subsequent works \cite{DeCalan:1972ya,Muta:1981pe,Lavelle:2005bt,Mirza:2006tk,Lavelle:2010hq} summed over some, but not all, possible degenerate initial and final states.  Ito \cite{Ito:1981nq} and Akhoury, Sotiropoulos, and Zakharov \cite{Akhoury:1997pb} included all possible disconnected amplitudes, including an infinite series of disconnected radiations, but, as was pointed out in \cite{Lavelle:2005bt}, the IASZ rearrangement of the formally divergent infinite series yields a NLO result that is exactly 0 at tree level.  

We report in this Letter a systematic treatment of Rutherford scattering at NLO in which we include all degenerate initial and final states.  We show for the first time how to rigorously perform the delicate rearrangement of the resultant formally divergent series.  The result is an IR safe NLO cross section whose small coupling limit is that from LO.  Although an alternative approach based on a coherent initial state exists \cite{Chung:1965zza,Kibble:1969kd,Curci:1978kj,Nelson:1980qs}, we find that the initial state soft degeneracy fully factorizes and cancels when using the usual particle physics wavepacket initial state.  Even though we only use our procedure in QED, we fully expect it to generalize to all other theories, thus explaining why all previous calculations that use the usual particle physics wavepacket initial state that neglected the soft initial state radiation degeneracy have nevertheless yielded cross sections in agreement with data.  

\section{NLO Rutherford}
\begin{figure}[!t]
    \subfloat[][]{
        \begin{fmffile}{LO}
            \begin{fmfgraph*}(90,70)
                \fmfleft{i1}
                \fmfright{o1}
                \fmfbottom{b1}
                \fmfleft{i2,i3}
                \fmfright{o2,o3}
                       \fmf{electron,lab.side=left,label=$p$}{i1,v1}
                       \fmf{electron,lab.side=left,label=$p^{\prime}$}{v1,o1}
                \fmf{photon,lab.side=left,label=$q$}{v1,b1}
                \fmfv{d.sh=cross,d.f=empty,d.si=.1w}{b1}
            \end{fmfgraph*}
        \end{fmffile}
        \label{subfig:LO}
    }
    \subfloat[][]{
        \begin{fmffile}{Vertex}
            \begin{fmfgraph*}(90,70)
                \fmfleft{i1}
                \fmfright{o1}
                \fmfbottom{b1}
                \fmfleft{i2,i3}
                \fmfright{o2,o3}
                       \fmf{electron,lab.side=left,label=$p$}{i1,v1}
                       \fmf{plain}{v1,v2}
                       \fmf{plain}{v2,v3}
                       \fmf{electron,lab.side=left,label=$p^{\prime}$}{v3,o1}
                \fmf{photon,left=0.6,tension=0.0,lab.side=left,label=$k$}{v1,v3}
                \fmf{photon,lab.side=left,label=$q$}{v2,b1}
                \fmfv{d.sh=cross,d.f=empty,d.si=.1w}{b1}
            \end{fmfgraph*}
        \end{fmffile}
        \label{subfig:Vertex}
    }\\
    \subfloat[][]{
        \begin{fmffile}{soft-em1}
            \begin{fmfgraph*}(90,70)
                \fmfleft{i1}
                \fmfright{o1}
                \fmfbottom{b1}
                \fmftop{t1,t2}
                \fmfleft{i2,i3}
                \fmfright{o2,o3}
                \fmf{photon,left=0.2,tension=0.0,lab.side=left,label=$k$}{v1,t2}
                \fmf{electron,lab.side=left,label=$p$}{i1,v1}
                \fmf{plain}{v1,v2}
                \fmf{plain}{v2,v3}
                \fmf{electron,lab.side=right,label=$p^{\prime}-k$}{v3,o1}
                \fmf{photon,lab.side=left,label=$q$}{v2,b1}
                \fmfv{d.sh=cross,d.f=empty,d.si=.1w}{b1}
            \end{fmfgraph*}
        \end{fmffile}
        \label{subfig:soft-em1}
    }
    \subfloat[][]{
        \begin{fmffile}{soft-em2}
            \begin{fmfgraph*}(90,70)
                \fmfleft{i1}
                \fmfright{o1}
                \fmfbottom{b1}
                \fmftop{t1,t2}
                \fmfleft{i2,i3}
                \fmfright{o2,o3}
                \fmf{photon,left=0.3,tension=0.0,lab.side=left,label=$k$}{v3,t2}
                \fmf{electron,lab.side=left,label=$p$}{i1,v1}
                \fmf{plain}{v1,v2}
                \fmf{plain}{v2,v3}
                \fmf{electron,lab.side=right,label=$p^{\prime}-k$}{v3,o1}
                \fmf{photon,lab.side=left,label=$q$}{v2,b1}
                \fmfv{d.sh=cross,d.f=empty,d.si=.1w}{b1}
            \end{fmfgraph*}
        \end{fmffile}
        \label{subfig:soft-em2}
    }\\
    \subfloat[][]{
        \begin{fmffile}{soft-abs1}
            \begin{fmfgraph*}(90,70)
                \fmfleft{i1}
                \fmfright{o1}
                \fmfbottom{b1}
                \fmftop{t1,t2}
                \fmfleft{i2,i3}
                \fmfright{o2,o3}
                \fmf{photon,right=0.3,tension=0.0,lab.side=right,label=$k$}{v1,t1}
                \fmf{electron,lab.side=right,label=$p-k$}{i1,v1}
                \fmf{plain}{v1,v2}
                \fmf{plain}{v2,v3}
                \fmf{electron,lab.side=left,label=$p^{\prime}$}{v3,o1}
                \fmf{photon,lab.side=left,label=$q$}{v2,b1}
                \fmfv{d.sh=cross,d.f=empty,d.si=.1w}{b1}
            \end{fmfgraph*}
        \end{fmffile}
    \label{subfig:initial-em1}
    }
    \subfloat[][]{
        \begin{fmffile}{soft-abs2}
            \begin{fmfgraph*}(90,70)
                \fmfleft{i1}
                \fmfright{o1}
                \fmfbottom{b1}
                \fmftop{t1,t2}
                \fmfleft{i2,i3}
                \fmfright{o2,o3}
                \fmf{photon,right=0.2,tension=0.0,lab.side=right,label=$k$}{v4,t1}
                \fmf{electron,lab.side=right,label=$p-k$}{i1,v2}
                \fmf{plain}{v2,v3}
                \fmf{plain}{v3,v4}
                \fmf{electron,lab.side=left,label=$p^{\prime}$}{v4,o1}
                \fmf{photon,lab.side=left,label=$q$}{v3,b1}
                \fmfv{d.sh=cross,d.f=empty,d.si=.1w}{b1}
            \end{fmfgraph*}
        \end{fmffile}
    \label{subfig:initial-em2}
    }
    \caption{The \protect\subref{subfig:LO} leading order (LO) and \protect\subref{subfig:Vertex} next-to-leading order (NLO) vertex contributions to the $1\rightarrow1$ process in Rutherford scattering. The \protect\subref{subfig:soft-em1} and \protect\subref{subfig:soft-em2} potentially degenerate final state radiative processes. The \protect\subref{subfig:initial-em1} and \protect\subref{subfig:initial-em2} potentially degenerate initial state radiative processes. }
    \label{fig:BN}
\end{figure}
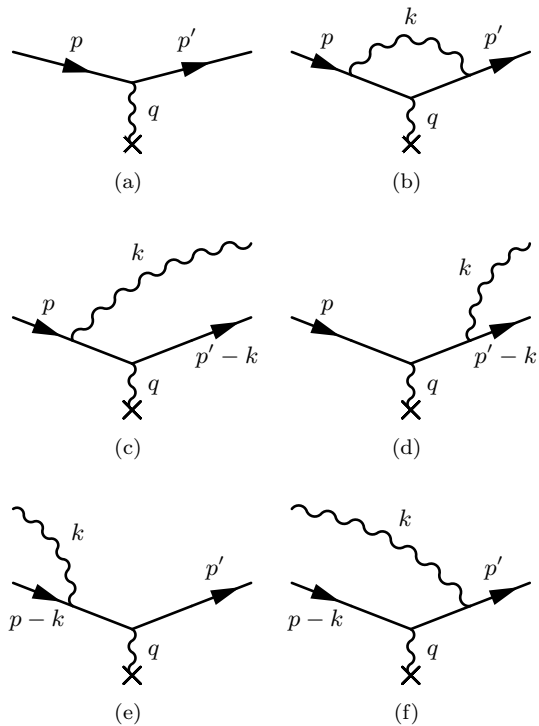

Consider an electron with inital four momentum $p=(E,\vec{p})$ that scatters off a static point charge (equivalently an external potential, as in LN \cite{Lee:1964is}) into a state with final four momentum $p'=(E,\vec{p}\,')$.  Define the momentum transfer $q\equiv p'-p$.  To regulate the various IR divergences, we introduce a fictitious photon mass $m_\gamma$ and keep the electron mass $m_e$ non-zero for now.  To apply the BN theorem to the NLO cross section we need to sum the LO \fig{BN} \subref{subfig:LO} and NLO vertex correction \subref{subfig:Vertex} for the $1\rightarrow1$ process as well as the indistinguishably soft final state radiative processes \subref{subfig:soft-em1} and \subref{subfig:soft-em2}.  
Summing these processes yields
\begin{multline}
    \label{eq:BN}
    d\sigma_{BN} 
    = d\sigma_0 \left\{1+\frac{\alpha_e}{\pi^2}\left[\log\left(\frac{E^2}{\Delta^2}\right)\left(1-\log\left(\frac{-q^2}{m_e^2}\right)\right) \right.\right. \\ \left.\left. +\frac{3}{2}\log\left(\frac{-q^2}{m_e^2}\right)+\mathcal{O}(1)\right]\right\},
\end{multline}
which is indeed free of soft IR divergences.  (In the $\overline{\mathrm{MS}}$ renormalization scheme used here, the $\log(-q^2/\mu_{\overline{\mathrm{MS}}}^2)$ terms from the vertex correction and the wavefunction renormalization exactly cancel.)

In order to remove the (potentially large) collinear logs, $\log(-q^2/m_e^2)$, we must invoke the KLN theorem and sum over the physically indistinguishable hard collinear final emission \fig{BN} \subref{subfig:soft-em2} and initial absorption \subref{subfig:initial-em1} processes.  (Soft collinear emission processes were already summed over in the application of the BN theorem.)  Notice that for the collinear divergence, requiring that the final electron three momentum $\vec{p}\,'$ is at a physically distinguishable angle $\phi>\delta$ from the incoming electron three momentum $\vec{p}$ means that these are the \emph{only} two additional diagrams we need to consider for indistinguishable hard collinear processes.  Adding these additional two processes to $d\sigma_{BN}$ yields a result in which one simply replaces $m_e^2$ with $\delta^2E^2$, which is free of all soft and collinear divergences and is the stopping point of the original LN paper \cite{Lee:1964is}.

Consistency with the principle of summing over all indistinguishable initial and final states, though, leads one to consider the degenerate soft initial photons (which includes photons that are both soft and collinear), \fig{BN} \subref{subfig:initial-em1} \emph{and} \subref{subfig:initial-em2}.  Including the contributions from the soft photons in these two diagrams, however, introduces new IR divergences, seemingly in contradiction of the KLN theorem.

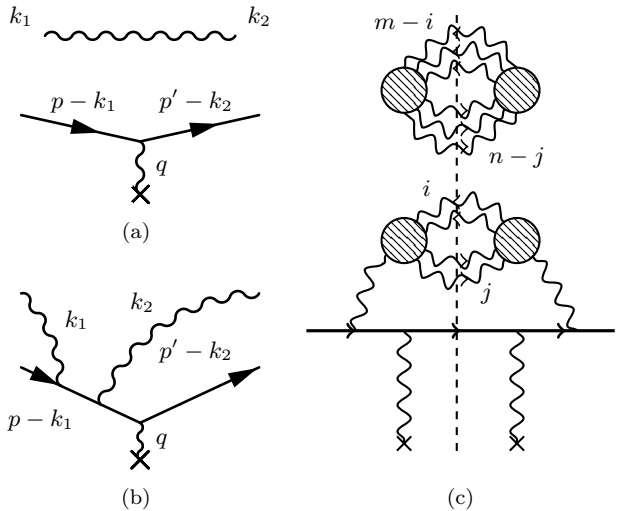
\begin{figure}[!htbp]
\begin{minipage}[b]{0.35\columnwidth}
    \subfloat[][]{
        \begin{fmffile}{LO-d}
            \begin{fmfgraph*}(90,60)
                \fmfleft{i1}
                \fmfright{o1}
                \fmfbottom{b1}
                \fmfleft{i2,i3}
                \fmfright{o2,o3}
                       \fmf{electron,lab.side=left,label=$p-k_1$}{i1,v1}
                       \fmf{electron,lab.side=left,label=$p^{\prime}-k_2$}{v1,o1}
                \fmf{photon,lab.side=left,label=$q$}{v1,b1}
                \fmf{photon}{i3,o3}
                \fmfv{label=$k_1$}{i3}
                \fmfv{label=$k_2$}{o3}
                \fmfv{d.sh=cross,d.f=empty,d.si=.1w}{b1}
            \end{fmfgraph*}
        \end{fmffile}
        \label{subfig:disconnected}
    }\\
    \subfloat[][]{
        \begin{fmffile}{soft-em-ab}
            \begin{fmfgraph*}(90,70)
                \fmfleft{i1}
                \fmfright{o1}
                \fmfbottom{b1}
                \fmftop{t1,t2}
                \fmfleft{i2,i3}
                \fmfright{o2,o3}
                \fmf{photon,right=0.2,tension=0.0,lab.side=right,label=$k_1$}{v1,t1}
                \fmf{photon,left=0.3,tension=0.0,lab.side=left,label=$k_2$}{v2,t2}
                \fmf{electron}{i1,v1}
                \fmf{plain,lab.side=right,label=$p-k_1$}{v1,v2}
                \fmf{plain}{v2,v3,v4,v5}
                \fmf{electron,lab.side=left,label=$p^{\prime}-k_2$}{v5,o1}
                \fmf{photon,lab.side=left,label=$q$}{v3,b1}
                \fmfv{d.sh=cross,d.f=empty,d.si=.1w}{b1}
            \end{fmfgraph*}
        \end{fmffile}
        \label{subfig:emissionabsorption}
    }
\end{minipage}
\begin{minipage}[t]{0.63\columnwidth}
    \hspace{0.2in}
    \subfloat[][]{
        \begin{tikzpicture}
            \draw[thick,pattern=north west lines] (0.0,3.2) circle [radius=0.3];
            \draw[thick,pattern=north west lines] (1.5,3.2) circle [radius=0.3];
            \draw[pho, thick] (1.34,2.95)..controls(0.75,2.6)..(0.16,2.95);
            \draw[pho, thick] (1.5,2.9)..controls(0.75,2.3)..(0.,2.9);
            \draw[pho, thick] (1.22,3.08)..controls(0.75,2.88)..(0.27,3.08);
            \draw[pho, thick] (0.28,3.3)..controls(0.73,3.55)..(1.22,3.3);
            \draw[pho, thick] (0.16,3.45)..controls(0.73,3.8)..(1.33,3.45);
            \draw[pho, thick] (0.,3.5)..controls(0.73,4.1)..(1.5,3.5);
            \node at (0,4.1) {$m-i$};
            \node at (1.5,2.3) {$n-j$};
            \draw[particle,very thick] (-1.3,0)--(0,0);
            \draw[particle,very thick] (0,0)--(1.5,0);
            \draw[particle,very thick](1.5,0)--(2.8,0);
            \draw[photon,thick] (0,0)--(0,-1.5);
            \draw[photon,thick] (1.5,0)--(1.5,-1.5);
            \draw[photon,thick] (-0.65,0)..controls(-0.45,0.8)..(-0.24,1);
            \draw[photon,thick] (2.3,0)..controls(2.,0.6)..(1.7,1);
            \draw[thick,pattern=north west lines] (0.0,1.2) circle [radius=0.3];
            \draw[thick,pattern=north west lines] (1.5,1.2) circle [radius=0.3];
            \draw[pho, thick] (1.34,0.95)..controls(0.75,0.6)..(0.16,0.95);
            \draw[pho, thick] (1.22,1.08)..controls(0.75,0.88)..(0.27,1.08);
            \draw[pho, thick,] (0.28,1.3)..controls(0.73,1.55)..(1.22,1.3);
            \draw[pho, thick] (0.16,1.45)..controls(0.73,1.8)..(1.33,1.45);
            \node at (0.3,1.9) {$i$};
            \node at (1.1,.5) {$j$};
            \newcommand{\Cross}{\mathbin{\tikz [x=1.4ex,y=1.4ex,line width=.2ex] \draw (0,0) -- (1,1) (0,1) -- (1,0);}}%
            \node at (0,-1.5) {$\Cross$};
            \node at (1.5,-1.5) {$\Cross$};
            \draw[dashed, thick] (0.7,4.2)--(0.7,-1.6);
        \end{tikzpicture}
        \label{subfig:cutdisconnected}
    }
\end{minipage}
    \caption{A disconnected photon amplitude such as \protect\subref{subfig:disconnected} might interfere with an amplitude with a photon absorption and emission process such as \protect\subref{subfig:emissionabsorption}.  \protect\subref{subfig:cutdisconnected} is a generic cut diagram with $m$ incoming and $n$ outgoing soft photons that is partially disconnected. (Note that the blobs include the possibility of connecting incoming photon lines from the left with conjugate incoming photon lines from the right.)
    }
    \label{fig:disconnected}
\end{figure}

What we will show is that the resolution is to include all possible contributions from degenerate initial and final states, including interference terms from partially disconnected amplitudes, i.e.\ amplitudes with disconnected sub-components.  \fig{disconnected} \subref{subfig:disconnected} is the simplest partially disconnected amplitude, and this partially disconnected amplitude yields a fully connected cut diagram when interfering with the fully connected amplitude \subref{subfig:emissionabsorption} that has both an absorption and emission process.  We must even include cut diagrams that have sub-components that are not connected to either the incoming or outgoing electron lines as shown in the generic cut diagram \fig{disconnected} \subref{subfig:cutdisconnected}.  

(Note that in these partially disconnected cut diagrams the incoming particle does scatter, $p'\ne p$, and therefore these cut diagrams \emph{are not} a part of the trivial 1 in the $S$ matrix.)

Following \cite{Ito:1981nq,Akhoury:1997pb}, we consider a general form of our process with $m$ incoming soft photons and $n$ outgoing soft photons:
\begin{align}
e^-+m \, \gamma \text{ (soft) } \rightarrow \text{ } e^-+n \, \gamma \text{ (soft)},
\label{4} 
\end{align}
with an amplitude $\mathcal{M}_{mn}$. Then the ``transition probability'' for the process becomes
\begin{align}
\mathcal{P}_{mn} = \frac{1}{m!}\frac{1}{n!} \sum_{i,f} \left|\mathcal{M}_{mn}\right|^2 ,
\label{5}
\end{align}
where $\mathcal{P}_{mn}$ contains contributions from both fully connected and partially disconnected cut diagrams and is summed over initial $i$ and final $f$ states. The total Lee-Nauenberg probability will be
\begin{align}
\mathcal{P} & = \sum_{m,n=0}^{\infty} \mathcal{P}_{mn} ,
\label{6}
\end{align}
where the KLN theorem ensures that the quantity $\mathcal{P}$ is free of soft or collinear IR divergences.

It is shown in \cite{Ito:1981nq,Akhoury:1997pb} that any cut diagram from $\mathcal{P}_{mn}$ at NLO can be constructed from four essential probabilities: $\mathcal{P}_{00}$, the cut diagram with no photons in the initial and final states (which may include the leading term, the vertex correction, the vacuum polarization, etc.); $\mathcal{P}_{10}$ and $\mathcal{P}_{01}$, which includes all cut diagrams with one soft photon in the initial state or final state, respectively; and $\widetilde{\mathcal{P}}_{11}$, which includes all fully connected cut diagrams with a single photon in both the initial state and final state. At NLO only one photon may attach to either the incoming or outgoing electron line, and the fully connected cut diagrams are given by one of the four previous basic probabilities.  Partially disconnected cut diagrams include the possibility of a photon connecting to the electron lines as well as any number of disconnected photon lines; these partially disconnected cut diagrams are given by one of the four probabilities multiplied by a number of $\delta$ functions according to the number of disconnected photons in the cut diagram. 

Define the function $\mathcal{D}(m-i,n-j)$ that describes the number of $m-i$ incoming and $n-j$ outgoing soft photons in $\mathcal{P}_{mn}$ disconnected from the electron lines. It is straightforward to see that $\mathcal{D}(0,0) = 1$ and $\mathcal{D}(a,b) = 0$ for $a\neq b$.  One may show that \cite{Ito:1981nq,Akhoury:1997pb}
the transition probability at NLO of $m$ incoming and $n$ outgoing soft photons is
\begin{multline}
    \mathcal{P}_{mn} = \frac{\mathcal{D}(m,n)}{m!\,n!}\, \mathcal{P}_{00}
    +\sum_{i=0}\frac{\mathcal{D}(m-i,n-i-1)}{(m-i)!\,(n-i-1)!}\, \mathcal{P}_{01}\\+
    \sum_{i=0}\frac{\mathcal{D}(m-i-1,n-i)}{(m-i-1)!\,(n-i)!}\, \mathcal{P}_{10}\\
    +\sum_{i=0}\frac{\mathcal{D}(m-i-1,n-i-1)}{(m-i-1)!\,(n-i-1)!}\, \widetilde{\mathcal{P}}_{11} .
    \label{eq:Pmndiv}
\end{multline}

Since $\mathcal{D}(0,0)=1$ for every $m$ and $n$ there is always a term unsuppressed by any factorials; hence $\mathcal{P} = \sum_{m,n}\mathcal{P}_{mn}$ formally diverges.  

In order to manipulate \eq{Pmndiv} in a controlled way, we introduce a convergence factor that becomes small for large $i$: we take
\begin{equation}
     \mathcal{D}(m-i,n-j) \;\; \rightarrow \;\; \mathcal{D}(m-i,n-j)e^{-(i+j)/\Lambda}
\end{equation}
with $\Lambda\gg1$.  Let us examine the partial sum of $\mathcal{P}_{mn}$ up to the emission of $N$ soft photons.  Since $\mathcal{D}(m,n)=0$ for $m\ne n$, we may simplify our manipulations by replacing the double sum over $m$ and $n$ with a single sum over $n$.  With the above convergence factor, we are guaranteed that $\mathcal{P}=\lim_{\Lambda\rightarrow\infinity}\lim_{N\rightarrow\infinity}\mathcal{P}_N(\Lambda)$ converges.

One may rearrange the partial sum of \eq{Pmndiv} up to $N$ soft photons to find
\begin{multline}
    \mathcal{P}_{N}(\Lambda) = \sum_{n=0}^N \frac{\mathcal{D}(n,n)}{(n!)^2}\,\left[\mathcal{P}_{00}+e^{-\frac{1}{\Lambda}}\mathcal{P}_{01}\right] + \sum_{n=1}^N \sum_{i=1}^n \\
     \frac{\mathcal{D}(n-i,n-i)}{[(n-i)!]^2}\, \left[e^{-\frac{2i+1}{\Lambda}}\mathcal{P}_{01}+e^{-\frac{2i-1}{\Lambda}}\mathcal{P}_{10}+e^{-\frac{2i}{\Lambda}}\widetilde{\mathcal{P}}_{11}\right], 
    \label{eq:Pmnconv}
\end{multline}
As it must, the above equation yields the same result as \eq{Pmndiv} as well as the rearranged series IASZ consider \cite{Ito:1981nq,Akhoury:1997pb}, when all series are given a similar convergence factor.

When the order of limit is switched, however, our result is very different from IASZ.  We will show that, like IASZ, the disconnected contribution completely factors out, allowing for an IR finite cross section; unlike IASZ, crucially, our result still has the $\mathcal{P}_{00}$ term.  Therefore our NLO expression correctly reduces to the LO expression as the coupling goes to 0.  

Swapping limits in an infinite series is a delicate procedure.  We are guaranteed from the Monotone Convergence Theorem that $P_N(\Lambda)$ converges to the same result independent of the order of limits taken should our partial sum 1) monotonically increase in $N$ for each $\Lambda$ and 2) monotonically increase in $\Lambda$ for each $N$ \cite{Knapp}.  

We will find it easier to prove that our rearranged sum satisfies these two properties if we exploit \cite{Muta:1981pe} the fact that 
\begin{equation}
    \mathcal{P}_{01} + \mathcal{P}_{10}=-\widetilde{\mathcal{P}}_{11},
    \label{eq:Psum}
\end{equation}
where the probability to emit a soft radiation given a scattering occurred is the usual
\begin{equation}
    \mathcal{P}_{01} = e^2|\mathcal{M}_0|^2\sum_{\mathrm{\lambda}}\left|\frac{p\cdot\epsilon_\lambda}{p\cdot k}-\frac{p'\cdot\epsilon_\lambda}{p'\cdot k}\right|^2.
\end{equation}
Then we have
\begin{multline}
    \mathcal{P}_{N}(\Lambda) = \sum_{n=0}^N \frac{\mathcal{D}(n,n)}{(n!)^2}\,\left[\mathcal{P}_{00}+e^{-\frac{1}{\Lambda}}\mathcal{P}_{01}\right] + \\
    \sum_{n=1}^N\sum_{i=1}^n \frac{\mathcal{D}(n-i,n-i)}{[(n-i)!]^2}\,2\mathcal{P}_{01}\, e^{-\frac{2i}{\Lambda}} \,\left[\cosh\left(\frac{1}{\Lambda}\right)-1\right].
    \raisetag{45pt}
    \label{eq:Pmnrearr}
\end{multline}

Since $\mathcal{P}_{00}$, $\mathcal{P}_{01}$, $\Lambda$, and $\mathcal{D}(n,n)$ are all strictly positive, \eq{Pmnrearr} clearly increases monotonically in $N$ for fixed $\Lambda$.  

To show that $\mathcal{P}_N(\Lambda)$ increases monotonically in $\Lambda$ for fixed $N$, we take the derivative with respect to $\Lambda$:
\begin{align}
    \frac{d\mathcal{P}_{N}(\Lambda)}{d\Lambda} 
    = \sum_{n=0}^N\frac{\mathcal{D}(n,n)}{(n!)^2}\left[\frac{1}{\Lambda^2} \, e^{-\frac{1}{\Lambda}}\,\mathcal{P}_{01}\right] + \mathcal{O}\big( \frac{1}{\Lambda^3} \big).
    \label{eq:derivative}
\end{align}
Although one finds that the higher order in $1/\Lambda$ correction term is negative, for any $N$ we can find a $\Lambda$ large enough such that the first term, which is strictly positive, dominates.

We have thus proved that we may exchange limits for our rearranged formula \eq{Pmnrearr}, and we may evaluate the $\Lambda\rightarrow\infinity$ limit first, yielding our main result:
\begin{equation}
    \mathcal{P} = (\mathcal{P}_{00} + \mathcal{P}_{01})\sum_{n=0}^\infinity \frac{\mathcal{D}(n,n)}{(n!)^2}.
    \label{eq:factorization} 
\end{equation}

We find that all the soft initial state physics of the infinite number of undetectably soft photons completely factorizes.  When the cross section is computed, one simply divides out by this unobserved infinity.  We have thus rendered all soft and collinear IR divergences harmless.

The final fully IR safe NLO Rutherford cross section including vacuum polarization \cite{Khalil} is
\begin{multline}
    d\sigma_{NLO} = d\sigma_0 \left\{1+\frac{\alpha_e}{\pi^2}\left[\log\left(\frac{E^2}{\Delta^2}\right)\left(1-\log\left(\frac{-q^2}{\delta^2 E^2}\right)\right) \right.\right. \\ \left.\left. +\frac{3}{2}\log\left(\frac{-q^2}{\delta^2 E^2}\right)+\frac{2}{3}\log\left(\frac{-q^2}{\mu^2_{\overline{\mathrm{MS}}}}\right)-\frac{28}{9}\right]\right\} \\ + \frac{\pi\alpha_e^3 E}{q^2} \left(\frac{1}{|\vec{q}|}-\frac{1}{|\vec{p}|}\right).
    \label{eq:sigmanlo}
\end{multline}
The Lorentz invariance breaking final term in \eq{sigmanlo} is from the ``box'' correction in which the electron interacts twice with the infinitely massive particle (recall that $d\sigma_0$ also, necessarily, breaks Lorentz invariance).

\section{Conclusions}
A self-consistent application of the KLN theorem requires a sum over all degenerate initial and final states to arrive at an IR safe cross section.  We derived the NLO correction to the Rutherford scattering cross section, including the full summation over all degenerate initial and final states.  This summation is formally divergent; after introducing a convergence factor, we proved that our rearrangement of this summation allows one to safely exchange taking the limit of the convergence factor to zero prior to the infinite sum limit.  After taking the convergence factor to zero, there was a complete factorization of the disconnected, soft initial state radiation and a complete cancellation of all other partially and fully connected soft initial state radiation effects.  Consistent with intuitive reasoning, the infinite factor from the soft initial state radiation will cancel in the physically observable cross section.  

We therefore arrived at the extremely nontrivial result that for NLO Rutherford scattering the summation over \emph{all} indistinguishable initial and final states is equivalent to the summation over \emph{only} the initial hard collinear and final soft, hard collinear, and soft and collinear degenerate states.  

Our result trivially extends to all NLO QED and QCD processes since the crucial IASZ \cite{Ito:1981nq,Akhoury:1997pb} insight to write the sum of all soft contributions in terms of four basic building blocks, \eq{Pmndiv}, will hold for any QED or QCD process at NLO accuracy, and the critical identity, \eq{Psum}, for the factorization, \eq{factorization}, after the rearrangement, \eq{Pmnrearr}, also holds for both QED and QCD \cite{Muta:1981pe}.  Presumably our work can be further generalized to all theories and orders.  


As was noted by Weinberg \cite{Weinberg:1995mt}, ``no one has given a complete demonstration that the sums of transition rates that are free of infrared divergences are the only ones that are experimentally measurable.''  We believe that our work here can be expanded in the future to provide such a proof.

\section*{Acknowledgements}
The authors thank the South African National Research Foundation and the SA-CERN consortium for their support. AK also thanks the African Institute for Mathematical Sciences (AIMS) and the University of Cape Town for their support.

\bibliography{main}
\end{document}